\documentclass[12pt]{article}
\usepackage{epsf}

\begin{document}

\bibliographystyle{unsrt}

\vspace*{-.6in}
\thispagestyle{empty}
\begin{flushright}
DAMTP R-98/20\\
hep-th/9804180
\end{flushright}
\baselineskip = 20pt

\vspace{.5in}
{\Large\bfseries
\begin{center}
BPS States of the Non--Abelian Born--Infeld Action
\end{center}}
\vspace{.4in}

\begin{center}
D. Brecher\footnote{email d.r.brecher@damtp.cam.ac.uk.}\\
\emph{D.A.M.T.P., University of Cambridge, Cambridge CB3 9EW, U.K.}
\end{center}
\vspace{1in}

\begin{center}
\textbf{Abstract}
\end{center}
\begin{quotation}
We argue that the trace structure of the non--abelian Born--Infeld
action can be fixed by demanding that the action be linearised by certain
energy--minimising BPS--like configurations.  It is shown how instantons
in D4-branes, $SU(2)$ monopoles and dyons in D3-branes, and vortices in
D2-branes are all BPS states of the action recently proposed by
Tseytlin.  All such configurations can be dealt with
exactly within the context of non--abelian Born--Infeld theory since, given
the relevant BPS--like condition, the action reduces to that
of Yang--Mills theory.  It would seem, moreover, that such an
analysis holds for the symmetrised trace structure of Tseytlin's
proposal only.
\end{quotation}
\vfil

\newpage

\pagenumbering{arabic}

\section{Introduction}

The Dirac--Born--Infeld (DBI) action has some remarkable properties.
Not least is the fact that it ``knows'' about energy--minimising BPS
states, or worldvolume ``solitons'', in the sense that it admits a
supersymmetric extension.  For such states, the action is linearised,
reducing to the simpler Maxwell theory.  The worldvolume theories of
D$p$-branes with specific values of $p$ admit the following such
solitons: abelian instantons in the D4-brane~\cite{gauntlett+etal};
abelian monoploes and dyons in the D3-brane~\cite{gibbons}; abelian
vortices in the D2-brane~\cite{callan+mal,gibbons}; and kinks in the
D-string~\cite{dasgupta+mukhi}.  For all these cases, it was shown
in~\cite{gauntlett+etal} that, given the relevant BPS--like condition,
the energy is minimised; in fact, all of the above configurations
follow from the D4-brane case by successive applications of
T--duality.

In this latter work it was, furthermore, claimed that the same ideas
should hold for the \emph{non}--abelian generalisation of the DBI
action, relevant to the description of multiple D-branes; that,
indeed, such properties could be viewed as criteria for fixing the
form of this, the non--abelian Born--Infeld (NBI) action.  This is the
view taken in
this letter, in which we consider the non--abelian generalisation of
the results of~\cite{gauntlett+etal} concentrating, where necessary,
on the $SU(2)$ case for definiteness.  Although there has been some
work on the question as to what is the correct generalisation of the
DBI action to the non--abelian case, the issue still seems to be
somewhat ambiguous.  We will show here that the action recently
proposed by
Tseytlin~\cite{tseytlin}, and verified in~\cite{me}, is singled out by
demanding such BPS properties; the arguments being really very simple.
This would suggest the existence of some supersymmetric extension of
this action, as oppposed to any other.

\section{General Considerations}

$SU(N)$ Yang--Mills theory should provide a good description of the
relevant dynamics of $N$ coincident D-branes~\cite{witten}.  It would
seem, however, that an NBI action, of which Yang--Mills theory is just
a ``non--relativistic'' approximation, should be used.  The natural
such action would be a generalisation of the DBI action, in which the
field strength is replaced by its non--abelian counterpart, and in
which the
worldvolume metric is multiplied by a unit matrix in the group space.
Then, since the action must be a \emph{group} scalar, we should trace
over it, e.g.~\cite{pol:tasi}:
\begin{equation}
{\cal L}_p = T_p ~{\rm Tr} \left[ I - \sqrt{-\det(\eta_{ab}I +
F_{ab})} \right],
\label{nbitr}
\end{equation}

\noindent where $I$ is the unit $SU(N)$ matrix, $F_{ab} = \partial_a
A_b - \partial_b A_a -i [A_a, A_b]= F_{ab}^i t^i$ is the non--abelian
field strength, $\{t^i, i=1,\ldots,N^2-1\}$ are an hermitian basis of
the $SU(N)$ algebra,
$[t^i, t^j] = i \varepsilon^{ijk} t^k$, and the trace is over the
fundamental representation.  We work throughout in units such that
$2\pi\alpha' =
1$; the tension of the branes is then $T_p = g_s^{-1}
(2\pi)^{(1-p)/2}$, which includes a
factor of the string coupling constant $g_s = e^{\Phi}$.  We have, for
the time being, ignored the (matrix--valued) transverse
coordinates\footnote{A note on
indices: $a,b = 0,1,\ldots,p$ denote worldvolume directions; $\alpha,
\beta = 1,\ldots,p$ denote world\emph{space} directions; $\mu, \nu =
p+1, \ldots, 9$ denote directions transverse to the brane; and $i,j$
will be group indices.}.  As explained in~\cite{me}, other trace
structures, such as $\sqrt{{\rm Tr}(-\det (\eta_{ab}I + F_{ab}))}$,
which is used implicitly in~\cite{gauntlett+etal}, can be ruled out
immediately.

We can, however, consider different group trace operations, and this
is where some ambiguity over the form of the action appears.  Tseytlin
has argued~\cite{tseytlin} that the NBI action should take the form
\begin{equation}
{\cal L}_p = T_p ~{\rm STr} \left[ I - \sqrt{-\det(\eta_{ab}I +
F_{ab})} \right],
\label{nbistr}
\end{equation}

\noindent where ${\rm STr}$ is a symmetrised trace, given by ${\rm
STr}(M_1,\ldots, M_n) = \frac{1}{n!} \sum_{\pi} {\rm Tr}(M_{\pi(1)}
\ldots M_{\pi(n)})$.  By including all possible permutations of
the matrices, the ${\rm STr}$ operation
resolves the matrix--ordering ambiguities involved in taking the
determinant of
a matrix--valued function.  We can also consider an antisymmetrised
trace ${\rm ATr}(M_1, \ldots, M_n) = \frac{1}{n!}\sum_{\pi}
(-1)^{\pi}{\rm
Tr}(M_{\pi(1)} \ldots M_{\pi(n)})$ and make use of the combination
${\rm STr} + i{\rm ATr}$,
e.g.~\cite{argyres+nappi}, the factor of $i$ being necessary since
the basis of the group algebra is hermitian.  These would seem to be
the only possibilities.  We will argue that, of these three trace
structures, Tseytlin's proposal is the only one which allows for the
BPS properties with which this paper is concerned. 

To this end, then, we will first consider D4-branes, with no scalar
fields excited; that is, we set $X^{\mu} = {\rm constant}$.  The
expansion of the
spacetime determinant in (\ref{nbistr}) gives
\begin{equation}
-\det(\eta_{ab} I + F_{ab}) = I + \frac{1}{2}F^2 + \frac{1}{3}F^3 -
\frac{1}{4}\left[F^4 - \frac{1}{2}(F^2)^2\right] + \frac{1}{5}F^5 +
\frac{1}{12}(F^2F^3 + F^3F^2),
\label{detexp4}
\end{equation}

\noindent where $F^2 = F_{ab}F^{ab}$, $F^3 = F_{ab}F^{bc}F_c^{~a}$,
$F^4 = F_{ab} F^{bc}F_{cd}F^{da}$ and
$F^5=F_{ab}F^{bc}F_{cd}F^{de}F_e^{~a}$.  In the abelian case, all the
odd powers of $F_{ab}$ vanish identically, but this is not so for the
case at hand.  A binomial expansion of this expression
results in a
infinite series, which is where the difficulties in the ${\rm
STr}$ prescription occur, since we must expand the binomial series
before the trace can be taken\footnote{Interesting progress has
recently been made~\cite{latest}, in which it has been shown that the
action describing two
D0-branes can be written in a closed form, after having taken the
symmetrised trace.}.
At any rate, it should be clear that the resulting expansion is just a
sum of both even and odd powers of $F_{ab}$.

The important properties of the ${\rm STr}$ and ${\rm ATr}$ operations
is that they pick
out the even and odd powers of $F_{ab}$
respectively.  Moreover, it is clear that, at least for the $SU(2)$
case, and to the first few orders, the same will apply to the cross
terms generated by the binomial expansion; that is, e.g., ${\rm STr}
(F^2 F^3) = {\rm STr}(F^2 F^5) = 0$.  So we can state, quite
generally, that the lagrangian (\ref{nbistr}) can be written as a sum
of even powers of $F_{ab}$ alone, whereas if we were to use either
{\rm Tr} or {\rm ATr}, odd powers would also be included.  Indeed,
this was the motivation behind Tseytlin's proposal.  Since odd
powers of $F_{ab}$ can be written in terms of derivatives of $F_{ab}$,
e.g. $F^3 \sim [F, F] F \sim [D, D]F$,
and in analogy with the abelian case, in which the DBI action does not
include derivatives of the field strength, Tseytlin was led simply to
define the NBI action to depend on even powers of $F_{ab}$ alone.

It must be noted that this discussion should be
viewed from within the context of a general analysis of possible NBI
actions.  That is, the action relevant to the description of multiple
D-branes has its origin in open superstring theory coupled to a
non--abelian gauge field; and it is known that the effective action of
this latter does not contain a term of the form
$F^3$~\cite{tseytlin2,me}.  From the point of view of string theory,
then, it would seem that the ${\rm STr}$ prescription alone is
acceptable.  Given that we want the NBI action to have the BPS properties
discussed, the arguments presented here should then be taken as evidence
for the ${\rm STr}$ prescription, string theory aside.

\section{Instantons in D4-Branes}

For static configurations of D4-branes, $F_{\alpha 0} = E_{\alpha} =
0$, and we have $-\det(\eta_{ab} I + F_{ab}) = \det(\delta_{\alpha\beta}
I + F_{\alpha\beta})$.  Then~\cite{gibbons+tseytlin,gauntlett+etal}
\begin{eqnarray}
{\cal L}_4 &=& T_4 ~{\rm STr} \left[ I - \sqrt{I + \frac{1}{4}F^2 +
\frac{1}{4}\tilde{F}^2  + \frac{1}{16} \left( F\cdot\tilde{F} \right)^2}
\right] \nonumber\\
 &=&
T_4~{\rm STr} \left[ I - \sqrt{ \left( I \pm
\frac{1}{4} F \cdot \tilde{F} \right)^2 -
\frac{1}{4} {\rm tr} \left| F \mp \tilde{F} \right|^2 }
\right],
\label{nbi4}
\end{eqnarray}

\noindent where, since we are dealing with static configurations,
$\tilde{F}_{\alpha\beta}$
is the Hodge dual of $F_{\alpha\beta}$, with respect to the world\emph{space}
indices only, and $F\cdot\tilde{F} =  F_{\alpha\beta}
\tilde{F}_{\alpha\beta}$.  It is important to note that in deriving
this equation, and those appearing below, use has been made of
the symmetry properties of the ${\rm STr}$ operation.  That is, the
matrices can
be treated as if they were \emph{abelian} until the last, at which point the
non--commuting group generators can be
re--inserted~\cite{hash+wash,hash}.  This seems somewhat
odd, but is in fact
not at all since, under ${\rm STr}$, we can assume $AB = BA$.
Any matrices can be freely interchanged under this operation;
and the procedure in~\cite{hash+wash,hash} is then fully justified.
Indeed, such a procedure automatically removes all odd powers of
$F_{ab}$ explicitly since, for the abelian case, all such powers
vanish identically.

It is easy to see, then, that for the (anti--)self--dual
configuration, for which $F_{\alpha\beta} = \pm \tilde{F}_{\alpha\beta}$,
the determinant can be written as a complete square, as has already
been noted
in~\cite{hash,me}.  The action is then linearised, becoming that of
Yang--Mills theory.  Since we are dealing with static configurations,
the energy density is just $T^{00}_4 = -{\cal L}_4$, this being
minimised if and only if
$F_{\alpha\beta} = \pm \tilde{F}_{\alpha\beta}$.  Thus, the
(anti--)self--duality condition at once linearises the action and
minimises the energy; hence its BPS interpretation.  We will see such
properties for all the static configurations considered below.  The
(anti--)self--duality condition
is, as usual, solved by multi--(anti\nolinebreak --) instanton
configurations, D0-branes from the worldvolume point of view.

How is this story changed if we were to use a different group trace
operation?  That is, how do the above considerations single out the
${\rm STr}$ operation alone?  A very simple argument shows that, if we
were to use either ${\rm Tr}$ or ${\rm STr} +i{\rm ATr}$,
the above analysis would no longer follow through.  In either of these
cases, odd powers of $F_{\alpha\beta}$ would be introduced into the
NBI action (\ref{nbi4}).  In the generic case, we would have to
consider the ${\cal O}(F^5)$
terms in (\ref{detexp4}) although, for the static case at hand, the
only additional term is that of $F^3 =
F_{\alpha\beta}F_{\beta\gamma}F_{\gamma\alpha}$.  Under either ${\rm
Tr}$ or ${\rm ATr}$, this term can be rewritten in terms of the dual
field strength as $F^3 = \tilde{F}_{\alpha\beta} \tilde{F}_{\beta\gamma}
F_{\gamma\alpha}$.  It is unclear, however, as to
how such a term could be included in the structure of (\ref{nbi4})
viz., a sum of squares.  Under the standard ${\rm Tr}$ operation, the
picture becomes more complicated still, since we can no longer
treat the matrices as if abelian.  At any rate, it
would seem that the determinant simply cannot
be written as a sum of squares if we are to include odd powers of
$F_{\alpha\beta}$.  If we cannot write the determinant as a sum of
squares, it certainly will not reduce to the linear Yang--Mills action
given the (anti--)self--duality condition.  It would be difficult to
claim that the energy is minimised by such configurations in this
case, since the NBI action cannot be linearised in any simple fashion.
If we are
to demand both that the action is linearised, and that the energy is
minimised by such (anti--)self--dual configurations, we simply cannot
include odd powers of $F_{\alpha\beta}$.  And the only way of ensuring
this is to use the ${\rm STr}$ operation and no other.  This same
argument holds for all the configurations considered below.

\section{Monopoles and Dyons in D3-Branes}

Since D0-branes in D4-branes are T--dual to D-strings ending on
D3-branes, the above instanton configurations will
give rise to monopoles and dyons within
D3-branes~\cite{gauntlett+etal}.  In~\cite{hash}, it was shown how the
standard $SU(2)$ BPS monopole solution
of Yang--Mills theory can be interpreted as a D-string joining
two D3-branes; and, via S--duality, a fundamental string joining the
two branes.  This work follows through for
precisely the reason that the BPS condition linearises the NBI action,
and so a mapping between the latter and the Yang--Mills action is
possible.  We will consider the dyonic generalisation of these results
here.

Including the transverse scalars, we have~\cite{tseytlin,me}
\[
{\cal L}_p = T_p ~{\rm STr} \left[ I - \sqrt{ \det( \delta_{\mu\nu} I
-i[X_{\mu},X_{\nu}] ) } \right.
\]
\begin{equation}
\times \left. \sqrt{ -\det( \eta_{ab} I + D_a X_{\mu} \left(
\delta_{\mu\nu} I
-i[X_{\mu}, X_{\nu}] \right)^{-1} D_b X_{\nu} + F_{ab} ) } \right].
\label{allx1}
\end{equation}

\noindent Since we will excite a single scalar only, the commutators
vanish identically here; such a simplification cannot be made,
however, in sections to follow.  We excite both the
electric and magnetic worldvolume
fields, and impose the ``static'' condition $D_0 X = 0$, in which
case the lagrangian is given by
\begin{equation}
{\cal L}_3 = T_3 ~{\rm STr} \left[ I - \sqrt{I + |\vec DX|^2 -|\vec
E|^2 +|\vec B|^2 - |\vec E \cdot \vec B|^2 - |\vec E \times \vec DX|^2
+ \left( \vec B \cdot \vec DX \right)^2} \right].
\end{equation}

\noindent The energy density is no longer simply the negative of the
lagrangian density, however; and this is where our results depart
somewhat from those of~\cite{gauntlett+etal}.  That is, the latter
work seems to make use of an energy functional different to that
of~\cite{callan+mal,gibbons}, for reasons which are unclear to the
author.

We claim here that the energy density can be given by a similar
Legendre transform as in the abelian case~\cite{gibbons+rasheed}.
Since taking the
variation of the determinant in the NBI action is an operation which
commutes with {\rm STr}, $\delta ({\rm STr} (F^n)) = {\rm
STr}(\delta(F^n))$, we can, as usual, define the
energy--momentum tensor by $T^{ab}_p \sim \frac{\delta {\cal
L}_p}{\delta g_{ab}}$.  The result is just what we find via the
prescription
\begin{equation}
T^{00}_p = \vec D^i \cdot \vec E^i - {\cal L}_p,
\label{energy}
\end{equation}

\noindent where $\vec D^i = \frac{\partial {\cal L}_p}{\partial \vec
E^i}$.  Using (\ref{energy}), we then have
\begin{equation}
T^{00}_3 = T_3 ~{\rm STr} \left[ \frac{(I + |\vec DX|^2)(I + |\vec
B|^2)}{\sqrt{I + |\vec DX|^2 -|\vec E|^2 +|\vec B|^2 - |\vec E \times
\vec DX|^2 - |\vec E \cdot \vec B|^2 + \left( \vec B \cdot \vec DX
\right) ^2 } }  - I
\right],
\end{equation}

\noindent where the denominator must be viewed formally as the inverse
of some matrix--valued (binomial) function.  There is no inconsistency
here since the fields can be taken to be abelian, as explained above.  At
any rate, we
see that the energy is bounded from below by configurations for which
$\vec E=\sin \theta \vec DX$ and $\vec B = \cos \theta \vec DX$, for
arbitrary angle $\theta$~\cite{gauntlett+etal}.  In this case, the
lagrangian is again linearised, and we have
\begin{equation}
{\cal L}_3 = -\frac{T_3(2\pi\alpha'g)^2}{2}~ {\rm Tr} \left[ |\vec B|^2
- |\vec E|^2 + (2\pi\alpha'g)^{-2}|\vec DX|^2 \right],
\label{nbi3}
\end{equation}
\begin{equation}
T^{00}_3 = \frac{T_3(2\pi\alpha'g)^2}{2} ~{\rm Tr} \left[ |\vec B|^2 +
|\vec E|^2 + (2\pi\alpha'g)^{-2}|\vec DX|^2 \right],
\end{equation}

\noindent where we have reinserted the relevant factors of
$2\pi\alpha'$ and the coupling constant $g$.  Comparing
with the usual Yang--Mills lagrangian, ${\cal L}_{YM} = -{\rm Tr}
\left[|\vec B|^2 - |\vec E|^2 + |\vec D \Phi|^2 \right]$, we see that
if we take $\frac{1}{2}T_3 (2\pi\alpha')^2 = g^{-2}$ and
$(2\pi\alpha'g)^{-1} X^i =
 \Phi^i$ solutions of the Yang--Mills theory will be solutions of the
linearised NBI theory~\cite{hash}.

The Prasad and Sommerfield solution~\cite{PS} is given in terms of the ansatz
\begin{equation}
\left. \begin{array}{ccl} A_{\alpha}^i & = &
\varepsilon_{i\alpha\beta} \hat{x}_{\beta} [1 - K(r)]/gr, \\ [.07in]
A_0^i & = & \hat{x}_i J(r)/gr, \\ [0.07in]
\Phi^i & = & \hat{x}_i H(r)/gr, \end{array} \right\}
\label{ansatz}
\end{equation}

\noindent where $r$ is a radial coordinate and $\{\hat{x}_i\}$ are unit
vectors.  The dyonic solution, satisfying $\vec E = \sin \theta \vec D
\Phi$ and $\vec B = \cos \theta \vec D \Phi$, has
\begin{equation}
\left. \begin{array}{ccl} K(r) & = & Cr/\sinh(Cr), \\ [0.07in]
J(r) & = & \frac{\sin \theta}{\cos \theta} [Cr \coth(Cr) -1 ], \\ [0.07in]
H(r) & = & \frac{1}{\cos \theta} [Cr \coth(Cr) -1 ], \end{array} \right\}
\end{equation}

\noindent where the constant $C = vg$, $v$ being the expectation value
of the Higgs field.  Standard analysis then gives $T^{00}_{YM} = v
\sqrt{q_E^2 + q_M^2}$ where $q_M = 4\pi/g$ and $q_E = \tan \theta
q_M$.
Asymptotically, the Higgs field can always be diagonalised by
performing a gauge transformation, interpreting
the diagonal entries as the asymptotic positions of the two branes: $X
= X^3 \sigma^3 / 2 = X(r)$.  Then
\begin{equation}
X(r) = \pm \frac{1}{2} \frac{(2\pi\alpha')}{\cos \theta} \left[ C
\coth(Cr) - \frac{1}{r} \right],
\label{bpshiggs}
\end{equation}

\noindent in which $\theta = 0$ corresponds to the results
of~\cite{hash}, the purely magnetic case.  Taking $C =
(2\pi\alpha')^{-1} \Delta X$, with
$\Delta X$
the separation of the branes gives, as
$r\rightarrow \infty$, $X(r) \rightarrow \pm (1 / \cos \theta) \Delta
X/2$.  Turning on the electric field (increasing $\theta$) increases
the effective separation of the
branes.  $\theta = \pi/2$, the purely electric case, gives an infinite
separation, corresponding to the semi--infinite string solution
of~\cite{callan+mal,gibbons}.  Quantum mechanically,
we can take $q_E = ng$, $n$ an integer.  Then, upon making the
relevant substitutions, we have
\begin{equation}
T^{00}_{YM} = (2\pi\alpha')^{-1} \Delta X \sqrt{ n^2 + \frac{1}{g_s^2}},
\end{equation}
\noindent which is precisely the energy of an $(n, 1)$ string of
length $\Delta X$, a dyon from the
worldvolume point of view~\cite{gauntlett+etal}.

\section{Vortices in D2-Branes and Hitchin's Equations}

By dimensionally reducing the (anti--)self--duality condition of
section three a second time, we obtain Hitchin's
equations~\cite{hitchin}: with $X$ and $Y$ the
two relevant transverse coordinates of the branes
\begin{equation}
\left. \begin{array}{ccl} \vec D X & = & \mp \star \vec D Y, \\ [0.07in]
\star F & = & \mp i [X, Y], \end{array} \right\}
\label{hitchin}
\end{equation}

\noindent which should describe non--abelian vortices in the
worldvolume of D2-branes~\cite{gauntlett+etal}.  As above, the Hodge
dual is taken with
respect to the spatial directions only: $\vec D = (D_1, D_2)$,
$\star \vec D = (D_2, -D_1)$ and $\star F = F_{12}$.

In the generic case, the two scalars will not be
simultaneously diagonalisable, the commutator terms in (\ref{allx1})
will be non--vanishing, and no natural interpretation in terms of
classical coordinates will be possible.  Then
\begin{equation}
\det ( \delta_{\mu\nu}I - i [X_{\mu}, X_{\nu}] ) = I - [X,Y]^2,
\end{equation}

\noindent and
\begin{equation}
D_a X_{\mu} ( \delta_{\mu\nu}I - i [X_{\mu}, X_{\nu}] )^{-1} D_b
X_{\nu} = ( I - [X,Y]^2 )^{-1} ( D_a X_{\mu} D_b X_{\mu} + i D_a X_{\mu}
[X_{\mu}, X_{\nu}] D_b X_{\nu} ).
\end{equation}

\noindent Evaluating the determinant over the $a,b$ indices
gives~\cite{gauntlett+etal}
\[
{\cal L}_2 = T_2 ~{\rm STr} \left[ I - \left\{ I + |\vec D X|^2 +
 |\star \vec D Y|^2 - [X,Y]^2 + |\star F|^2 \right.\right.
\]
\[
\left.\left. + |\vec D X \cdot \star \vec D Y|^2 - |\star F|^2 [X,Y]^2
+ 2i \star F (\vec D X \cdot \star \vec D Y) [X,Y] \right\}^{1/2} \right] =  
\]
\begin{equation}
 T_2 ~{\rm STr} \left[ I - \sqrt{ \left( I \mp ( \star F i[X,Y]
+ \vec D X \cdot \star \vec D Y) \right)^2 + |\vec D X \pm \star \vec
D Y|^2 +
|\star F \pm i [X,Y]|^2} \right].
\label{nbi2}
\end{equation}

\noindent Thus, the energy density $T^{00}_2 = - {\cal L}_2$ is
\begin{equation}
T^{00}_2 \geq \mp T_2~{\rm Tr} \left[ \star F i[X,Y] + \vec D X \cdot
\star \vec
D Y \right] = T_2~{\rm Tr} \left[ \frac{1}{2} |\vec D X|^2 +
\frac{1}{2} |\vec D
Y|^2 - [X,Y]^2 \right],
\end{equation}

\noindent the last step being valid for the energy--minimising
configurations for which Hitchin's equations (\ref{hitchin}) are
obeyed.  In this case, the lagrangian
(\ref{nbi2}) is once again linearised.  The, superficially
complicated, lagrangian (\ref{allx1}), involving the product of two
determinants taken over different indices reduces to the simple
dimensionally reduced Yang--Mills lagrangian, when the BPS--like
conditions hold.

\section{D-strings and Nahm's equations}

Nahm's equations~\cite{nahm}, the dimensionally reduced version of Hitchin's
equations, reduce the $SU(2)$ monopole problem in three (spatial)
dimensions to
a one--dimensional problem.  That is, the dyon of section four can be
described from within either the D3-branes' worldvolume, or from
within that of the string~\cite{gauntlett+etal}.  Indeed, it was shown
in~\cite{diaconescu} from the point of
view of Yang--Mills theory, that
Nahm's equations do in fact follow from the worldvolume theory of the
D-string.  Here we will show this to be true for the full--blown NBI
action.

To this end, we consider static configurations, in which case $F_{ab}
\rightarrow F_{10} =
E_1 = 0$.  Then the relevant lagrangian is the non--abelian
generalisation of the Dirac lagrangian, in which three scalars are excited:
\begin{equation}
{\cal L}_1 = T_1~{\rm STr} \left[ I - \sqrt{ \det ( \delta_{\mu\nu}I -
i[X_{\mu}, X_{\nu}] ) \det ( I + D X_{\mu} ( \delta_{\mu\nu}I -
i[X_{\mu}, X_{\nu}] )^{-1} D X_{\nu} ) } \right],
\label{nbi1}
\end{equation}

\noindent where $D_{\alpha} X_{\mu} \rightarrow D_1 X_{\mu} \equiv D
X_{\mu}$ and $\mu, \nu$ run over the three spatial directions of the
D3-brane.  Then
\begin{equation}
\det ( \delta_{\mu\nu}I - i[X_{\mu}, X_{\nu}] ) = I - \frac{1}{2} [
X_{\mu}, X_{\mu} ]^2,
\end{equation}

\noindent and
\begin{equation}
D X_{\mu} ( \delta_{\mu\nu}I - i[X_{\mu}, X_{\nu}] )^{-1} D X_{\nu} =
\left( I - \frac{1}{2} [ X_{\mu}, X_{\mu} ]^2 \right)^{-1} \left[ D
X_{\mu} D X_{\mu} - \left( \frac{1}{2} \varepsilon_{\mu\nu\rho}
X_{\mu} [ X_{\nu},
X_{\rho} ] \right)^2 \right].
\end{equation}

\noindent Substituting these latter into (\ref{nbi1}) gives the lagrangian
\begin{eqnarray}
{\cal L}_1 &=& T_1~{\rm STr} \left[ I - \sqrt{ I + D X_{\mu} D X_{\mu} -
\frac{1}{2} [X_{\mu}, X_{\nu}]^2 - \left( \frac{1}{2}
\varepsilon_{\mu\nu\rho} D X_{\mu} [ X_{\nu}, X_{\rho}] \right)^2 }
\right] \nonumber\\
&=&
T_1~{\rm STr} \left[ I - \sqrt{ \left( I \mp \frac{i}{2}
\varepsilon_{\mu\nu\rho} D X_{\mu} [X_{\nu}, X_{\rho}] \right)^2 +
{\rm tr} \left| D
X_{\mu} \mp \frac{i}{2} \varepsilon_{\mu\nu\rho} [X_{\nu},
X_{\rho}] \right|^2 } \right].
\end{eqnarray}

\noindent The energy density, $T^{00}_1 = - {\cal L}_1$, and so
\begin{equation}
T^{00}_1 \geq \pm T_1~\frac{i}{2} {\rm Tr} \left[
\varepsilon_{\mu\nu\rho} D X_{\mu}
[X_{\nu}, X_{\rho}] \right] = T_1~\frac{1}{2}{\rm Tr} \left[ D X_{\mu}
D X_{\mu}
- \frac{1}{2} [ X_{\mu}, X_{\nu} ]^2 \right],
\end{equation}

\noindent the last step being valid if Nahm's equations,
\begin{equation}
D X_{\mu} = \pm \frac{i}{2} \varepsilon_{\mu\nu\rho} [ X_{\nu},
X_{\rho}],
\end{equation}

\noindent are obeyed.  In this case, as expected, the action is
linearised, and the
energy minimised.

A final T--duality takes us back to the configuration of section
three, but now from the point of view of the D0-branes lying within
the D4-branes.  We excite the four relevant
scalars, those corresponding to the four spatial directions of the
D4-brane, and consider static configurations, in which case the
determinant over the $a,b$ indices in (\ref{allx1}) drops out
entirely.  As should be expected, the lagrangian is formally identical to
(\ref{nbi4}), with $F_{\alpha\beta} \rightarrow -i[X_{\mu}, X_{\nu}]$.
The energy $T^{00}_0 = -{\cal L}_0$ is minimised, and the lagrangian
is linearised if
\begin{equation}
[X_{\mu}, X_{\nu}] = \pm \varepsilon_{\mu\nu\rho\sigma} X_{\rho} X_{\sigma},
\end{equation}

\noindent which is just the (anti--)self--duality condition of section
three from the D0-branes' point of view.  When this condition
holds~\cite{gauntlett+etal}
\begin{equation}
T^{00}_0 = \mp T_0 ~\frac{1}{4}{\rm Tr} \left[
\varepsilon_{\mu\nu\rho\sigma} X_{\mu} X_{\nu} X_{\rho} X_{\sigma} \right],
\label{energy0}
\end{equation}

\noindent which vanishes identically for finite dimensionally
matrices.  In the M(atrix) theory $N \rightarrow \infty$ limit,
however, the energy density (\ref{energy0}) in the case of D0-branes
on $T^4$ corresponds to a single unit of D4-brane
charge~\cite{wash:review}.  In this limit, then, the instanton
configurations of section three can be described from the point of
view of the D0-branes'.

\section{Discussion}

The moral of the story is that we can use the facts that BPS--like
conditions should, firstly, linearise the NBI action and, secondly, should
minimise the NBI energy, as criteria to fix the trace structure of
this action.  That is, from the three possible trace structures which
such an action could have --- ${\rm Tr}$, ${\rm STr}$, and ${\rm STr} +
i {\rm ATr}$ --- Tseytlin's ${\rm STr}$ prescription is singled out.  In
this case, we have shown how certain worldvolume ``solitons'' are BPS
states of the NBI action.  All such configurations minimise the
worldvolume energy, this being due to the fact that, given the
relevant BPS--like condition, the determinant in
the NBI action can be written as a complete square.  The relevant
action for all these configurations is just that of (dimensionally
reduced) Yang--Mills theory.
The reason that ${\rm STr}$ is singled out by this analysis is
simple: only in this case do no odd powers of the field strength, or of
the worldvolume scalars, appear
in the NBI action and so, only in this case, can the determinants be
written as complete squares.  Only for this presciption, then, can
such exact statements concerning the form of the NBI action be made.
Moreover, as the usual BPS equation follows ultimately from supersymmetry
considerations, it would be natural to postulate that the ${\rm STr}$
prescription, and no other, allows a supersymmetric extension.\newline

\noindent {\bf Acknowledgments}

I would like to thank Malcolm Perry for a first reading of the draft
of this manuscript.

\end{document}